\documentclass[12pt]{iopart}

\usepackage{graphicx}
\usepackage{hyperref}
\usepackage{cleveref}
\usepackage{textcomp}
\usepackage[caption=false]{subfig}
\usepackage{tabularray}
\usepackage{xcolor}

\newcommand{\commonLongTitle}{Thermal Shrinkage-Induced Modifications in Photonic Band Gaps of Two-Photon Polymerized Bragg Reflectors}
\newcommand{\commonShortTitle}{Thermal Shrinkage-Induced Modifications in PBGs of 2PP Bragg Reflectors}

\makeatletter
\g@addto@macro\@floatboxreset\centering
\makeatother

\begin{document}

\title[\commonShortTitle]{\commonLongTitle}

\author{Yu-Shao Jacky Chen$^1$, Mike P. C. Taverne$^{1, 2}$, Kevin Chung-Che Huang$^3$, Ying-Lung Daniel Ho$^{1, 2}$ and John G. Rarity$^1$}

\address{$^1$ Department of Electrical and Electronic Engineering, University of Bristol, Bristol, UK}
\address{$^2$ Department of Mathematics, Physics \& Electrical Engineering, Northumbria University, Newcastle upon Tyne, UK}
\address{$^3$ Optoelectronics Research Centre, University of Southampton, Southampton, UK}

\ead{mike.taverne@northumbria.ac.uk, daniel.ho@northumbria.ac.uk, John.Rarity@bristol.ac.uk}

\begin{abstract}
One-dimensional (1D) polymer-based photonic crystals (PhCs) in the 1.55 µm wavelength range can be easily created using a two-photon direct laser writing system. 
To achieve shorter period structures, we report the use of thermal shrinkage of two-photon polymerized structures, at elevated temperatures, to eliminate unpolymerised material, leading to the uniform shrinkage of the distributed Bragg reflector (DBR) structures by a ratio of $\sim$ 2.5 to 5. 
Our Finite Difference Time Domain (FDTD) simulation and the angle-resolved light scattering characterization technique using Fourier image spectroscopy (FIS) measurements show that the low order photonic bandgaps of DBRs blue-shift across the NIR-visible region (850 to 400 nm) as the shrinkage increases.
\end{abstract}

\maketitle


\section{Introduction}
    Photonic crystals (PhC) were first suggested as candidates for the control of light by Sajeev John\cite{10.1103/PhysRevLett.58.2486} and Eli Yablonovitch in 1987\cite{10.1103/PhysRevLett.58.2059}.
    The concept of photonic bandgaps (PBGs) in the photonic crystal is similar to that of electronic bandgaps in the semiconductor, wherein the former refers to the forbidden region for photons and the latter for the electrons.
    In PBG materials, photons act like carriers like electrons in semiconducting materials. 
    Hence in the bandgaps the PBG material can prohibit light propagation in all directions and achieve low-loss reflection of light from all directions within a specific wavelength range. 
    Thus, a high quality factor resonant cavity strongly localising photons can be formed at crystal defects \cite{10.1038/nphoton.2007.141,10.1364/OL.43.005202,10.3390/cryst12030303,10.3390/app8071087,10.1209/0295-5075/116/64007,10.1364/JOSAB.32.000639,10.1109/JQE.2011.2170404}.
    Recently, researchers have demonstrated precise crystal design by the direct laser writing (DLW) system using two-photon lithography (2PP), which makes the fabrication of PhC becomes more feasible\cite{10.1002/adma.200600769, 10.3390/nano8070498, SPIE.2017}. 
    To date, numerous 3D PhC structures showing partial and complete PBGs have been designed and fabricated using DLW.\cite{10.1364/OL.42.001584, 10.1364/OE.23.026565, 10.1021/acsphotonics.9b00184,10.1021/acsaom.3c00055}. 
    However, the spatial resolution of DLW has been limited by the laser energy power distribution due to its Abbe diffraction limit\cite{10.1002/lpor.201100046}. 
    So, designing a PBG working at the visible range has been more challenging.
    The stimulated emission depletion direct laser writing (STED-DLW) technique has been applied in order to scale to sub-10 nanometer feature size \cite{10.1088/2399-1984/aabb94,10.1007/s12274-017-1694-0,10.3390/ma13030761}, 
    but challenges in maintaining the complex optical system alignment in STED-DLW have restricted its wide usage and commercialisation.

    In this paper, the band-gap shifts in distributed Bragg reflectors (DBRs) using the thermal shrinkage of two-photon polymerised structures are presented \cite{10.1016/j.mee.2018.01.018, 10.1038/s41378-019-0079-9, Liu2019_10.1038/s41467-019-12360-w, 10.3390/ma13030761}.
    This thermal shrinkage technique has been confirmed to uniformly shrink the woodpile-structured PhCs and successfully achieved band-gap tuning in the visible range. 
    Using this method, researchers can easily modify the colour of 3D PhC structures from white to purple \cite{Liu2019_10.1038/s41467-019-12360-w}.
    The polymer-based DBRs with pillar-supported layers \cite{10.1364/OL.43.004711} have been fabricated by Nanoscribe’s Dip-in Laser Lithography (DiLL) system, followed by thermal shrinkage of these polymerised structures from room temperature to elevated temperature up to 450 °C. 
    The broadband and wide angle-resolved scattering characterisation of DBRs in the visible and infrared range have been visualised by our in-house built Fourier image spectroscopy (FIS) system \cite{10.1364/OL.42.001584, 10.1021/acsphotonics.9b00184, 10.1364/OE.23.026565}. 
    A significant blue shift of the DBR's low order band gaps into the visible region has been demonstrated.

    


      
\section{Design and fabrication}
    \subsection{Template design and direct laser writing}
    \begin{figure}[b]
        \includegraphics[width=0.7\linewidth]{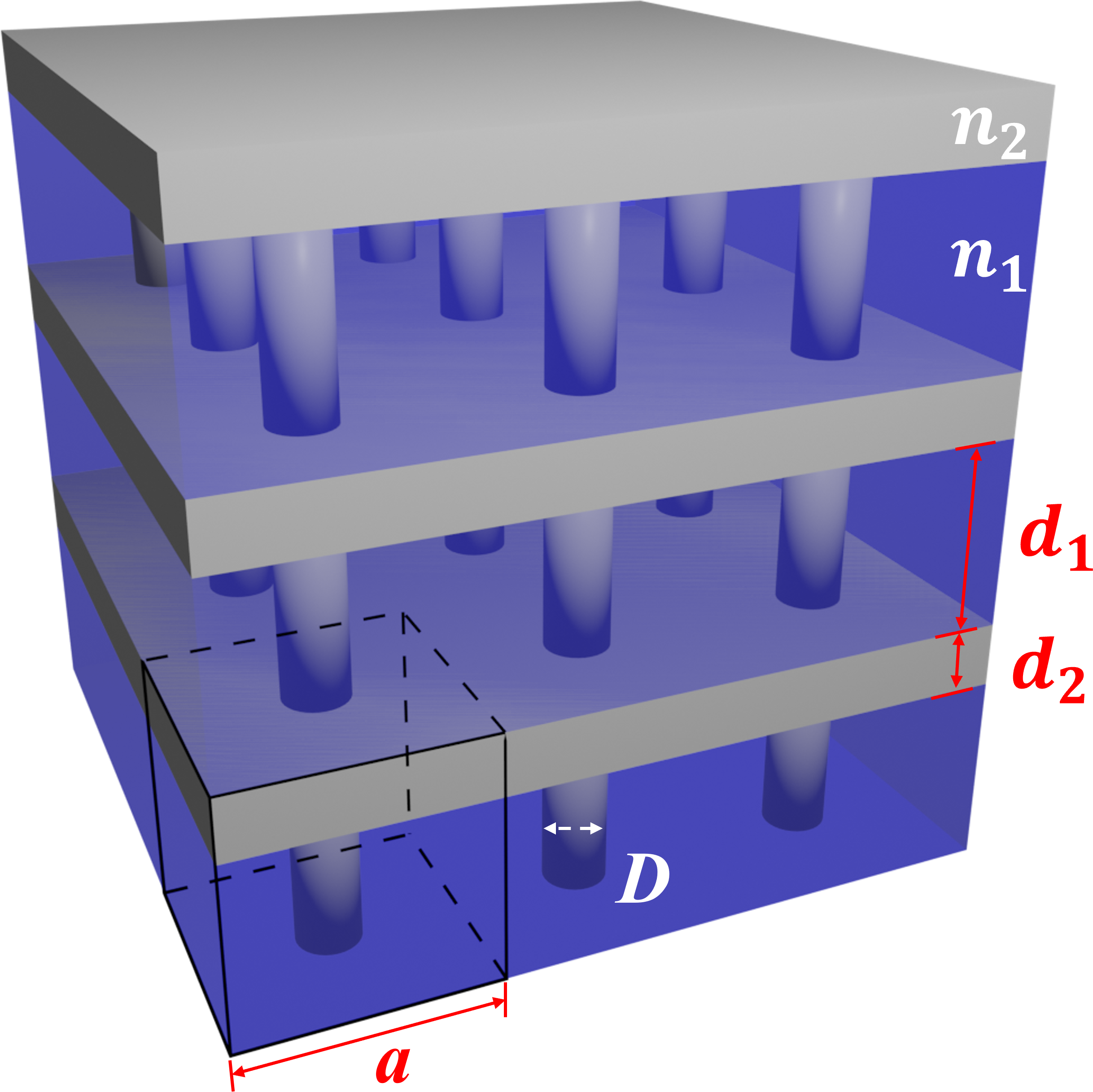}
        \caption{The illustrated 3x3x3 pillar supported DBR PhC, with the unit cell selected at the corner, \emph{a} is the unit cell size in the x-y period and the pillar period. \emph{D} is the pillar diameter; \emph{$d_1, d_2$} and \emph{$n_1,n_2$} refer to the thickness of the layer and the material refractive index, respectively.}
        \label{fig:illustrate}
\end{figure}

    DBRs are 1D PhCs that consist of sequentially arranged layers with varying refractive indices. These straightforward structured multilayers are designed to selectively reflect certain wavelengths while allowing the transmission of others, as illustrated in Fig.~\ref{fig:illustrate}.
    The said configuration is characterized by a unit-cell size of \(a\), pillar diameter of \(D\), air thickness of \(d_1\), polymer thickness of \(d_2\), as well as their corresponding refractive indices of \(n_1\) and \(n_2\).
    The polymer layers along the z-axis represent the pillar supports, which were designed to support the layers across the x and y axes of the structure. 
    Owing to their inherent geometric simplicity, DBRs offer ease in both fabrication and subsequent analysis. 
    The reflective characteristics of DBRs are attributed to the constructive interference of light waves at the interfaces between these alternating layers. 
    The specific wavelength at which reflection is maximised is governed by the individual layer thicknesses and the refractive index contrast between them.  
    By adjusting these parameters, DBRs can be tailored to reflect a desired wavelength range. However, conventional DBRs are typically fabricated by depositing alternating materials onto substrates, a method not compatible with the 2PP-DLW technique. Furthermore, in the context of 2PP-DLW based DBR fabrication, the choice of layer materials is predominantly confined to air and photoresist combinations.
    Considering the research objectives of this study, the design of the sample structure holds significance. 
    DLW has the potential to assist in designing various sample structures. 
    However, due to the limitations of the mechanism\cite{10.1063/5.0166905}, achieving a minimum resolution suitable for the visible range is challenging. As a result, the post-processing methodology becomes one of the candidates for achieving band structure modifications. 
    Based on the scaling property of the 3D PhC, we know the band gap is related to the unit cell scale. 
    Therefore, we aim to achieve the band modification by scaling down the desired structure.
    The most common 3D PhCs fabricated by DLW are made in polymer. Recently, thermal annealing leading to evaporation of residual monomer has gained attention due to its favourable shrinkage rate and a broader range of post-processing possibilities \cite{Liu2019_10.1038/s41467-019-12360-w,10.1016/j.mee.2018.01.018,10.1038/s41378-019-0079-9,10.3390/nano8070498,10.1002/adom.202200232}.  
    As a 1D PhC, the DBR possesses a well-defined photon bandstructure and relatively straightforward fabrication. By 2PP-DLW technique, it becomes feasible to fabricate complex and customised 3D geometry structures with ease. Here, the Nanoscribe's DiLL is utilized to create the DBR templates. In the DiLL process, the objective lens (63x Zeiss \cite{Nanoscribe_User_Manual_2017-10-23}) of the printing system is immersed directly into the photoresist (Ip-Dip\cite{Nanoscribe_User_Manual_2017-10-23}).
    Concerning the upcoming thermal post-processing, we opted for sturdier quartz substrates instead of the usual thin coverslips. These quartz cover slips are too thick to allow the standard oil immersion writing emphasising the need for the DiLL approach, where the polymer acts as its own 'oil immersion'. A femtosecond laser with 780 nm wavelength and power of 28 mW is then focused into the photoresist.
   

\begin{figure}
\subfloat[Top-view of DBR template][Top-view of DBR template]{
\includegraphics[width=0.7\linewidth]{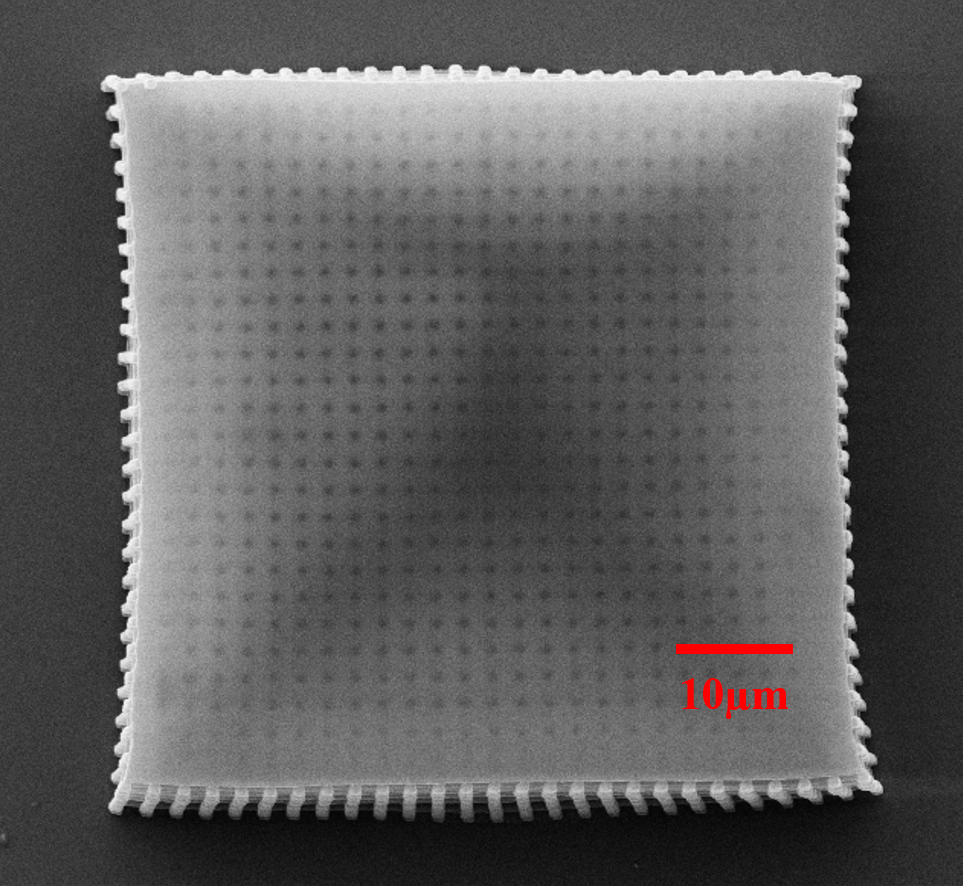}
\label{fig:Top-view}}

\subfloat[cross-section of DBR template][cross-section of DBR template]{
\includegraphics[width=0.7\linewidth]{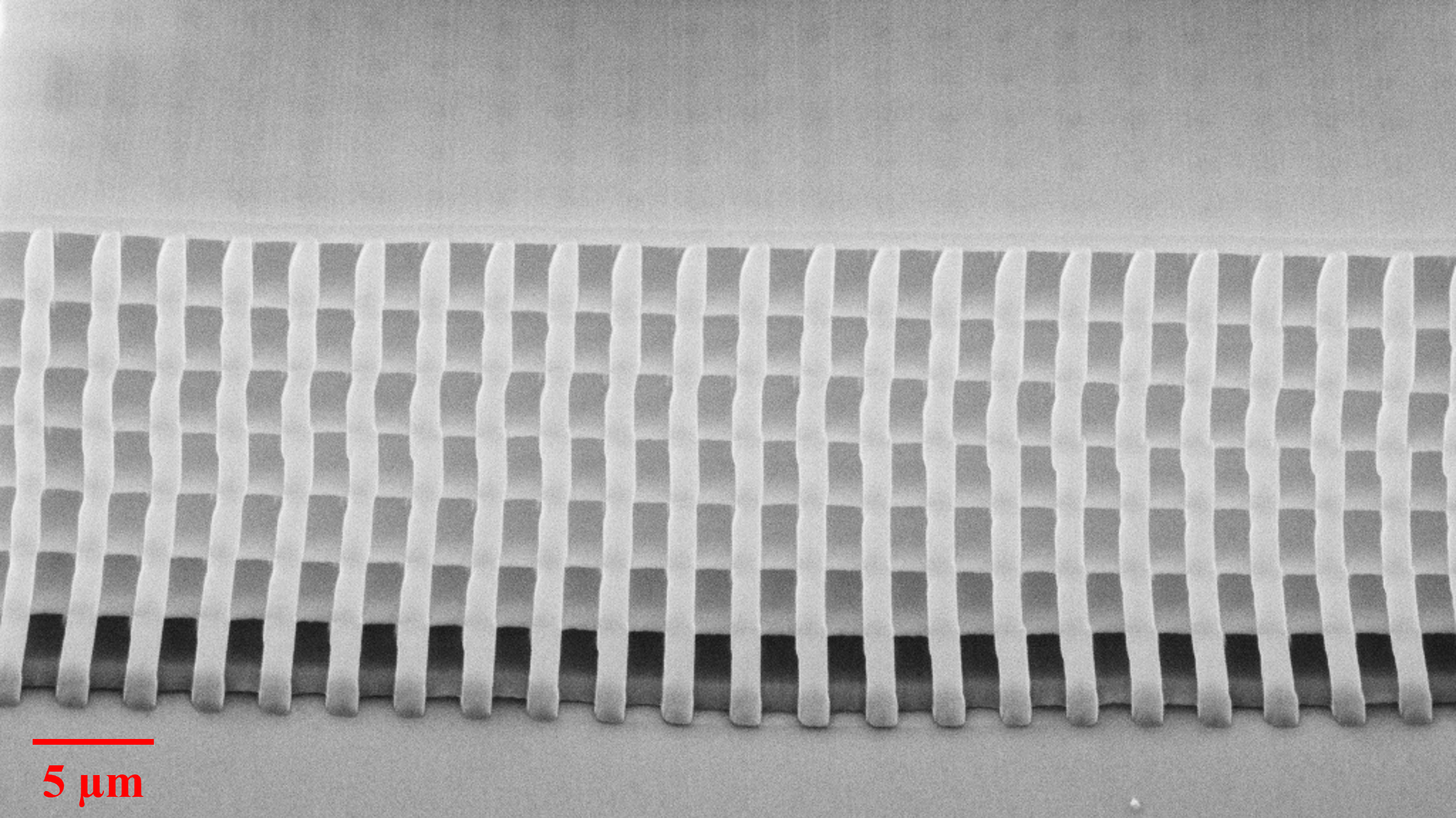}
\label{fig:cross-section}}

\caption{The SEM image of pillar-supported DBR demonstrated for this research. The \(d_1\) and \(d_2\) are the thickness of air and polymer. In this study, \(d_1\) is around 2.8µm and \(d_2\) is around 1µm. The detail of the parameter is shown in Table~\ref{table:parameter}. The actual templates for post-processing obtain more periods.}
\label{fig:SEM}
\end{figure}

    By controlling the laser's focus position in three dimensions, intricate 3D structures can be written with 100nm precision.  When the laser power is set for 19.5\% of full power (approximately 3.1-5.3 mW), restricted by an acoustic-optical modulator, this ensures the voxel depth and resulting polymer layer thickness ($d_2$) is close to 1 $\mu m$.
    The unpolymerized photoresist is subsequently washed away by the IPA, leaving behind the desired 3D microstructure, the demonstrated sample is shown in Fig.~\ref{fig:Top-view} and \ref{fig:cross-section}.
    
    \subsection{Thermal post-processing and shrinkage ratio}

    The polymerised DBRs are then post-processed using the Jipelec JetFirst rapid thermal annealing (RTA) system to scale down the structures and achieve the desired bandgap modification. 
    The thermal treatment vaporizes the solvent residue and can further transform the polymer into carbon when the temperature reaches a sufficient level\cite{10.1016/j.mee.2018.01.018,10.1038/s41378-019-0079-9}.
    Here, we have adopted the thermal annealing recipes reported by Liu et al.\cite{Liu2019_10.1038/s41467-019-12360-w}, followed by the optimal adjustment of parameters to create stable periodic structures.
    
\begin{figure}
    \includegraphics[width=0.8\linewidth]{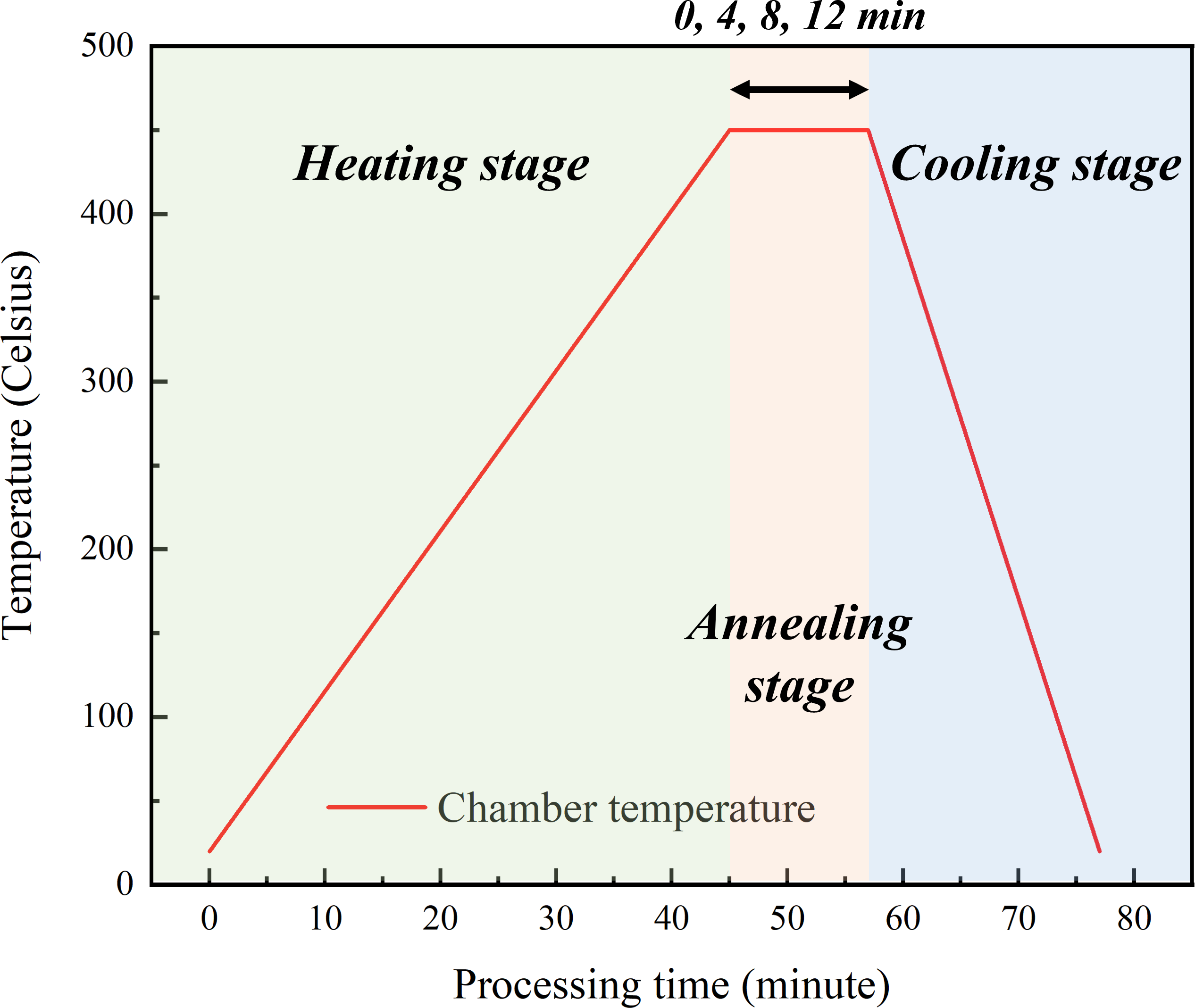}
    \caption{ Chamber temperature versus processing time for different annealing recipes. In each case, the temperature is first increased to a maximum of 450 ℃ with a heating rate of 10 ℃/min, then remains there for different times depending on the annealing recipe and finally cools back to room temperature at a rate of -20 ℃/min.}
    \label{fig:recipe}
\end{figure}

    The thermal post-processing recipe involves three main steps. Firstly, in the heating stage, the temperature is gradually increased up to 450~°C at a ramp rate of 10~°C/min.
    In the following annealing stage, the temperature is kept constant at 450~°C for an arbitrary time, which is the main variable in this post-processing.
    Finally, it ends with the cooling stage, where the chamber was cooled to room temperature over a span of $\sim$20 minutes. The shrinkage starts between 300 and 400~°C.
    A graphical representation of the temperature-time correlation is depicted in Fig.~\ref{fig:recipe}.
    
    At 450~℃, the templates undergo significant deformation, marking this temperature as the annealing point and a pivotal period for controlling shrinkage amount.
    This entire procedure is conducted under an argon flow of 50~sccm to prevent oxidation.
    However, the thermal-induced shrinkage deforms the layer, leading to the tilt on the top of structures. 
    With the inspiration from a previous work\cite{10.1002/adom.202200232, 10.1016/j.mee.2018.01.018}, 
    the first layer between the polymer template and substrate consists of a slightly longer pillar array. 
    These stretchable legs provide stress relief effectively reducing the geometric deformation after thermal shrinkage.
    Thus, the pillar arrays between the layers not only act as a support to prevent sagging of the template but also reduce the adhesion effects between the silica substrate and polymer template\cite{10.1063/5.0166905,10.1002/adom.202200232}. In early results we also saw stretch distortion of significant numbers of lower DBR layers. To ensure a significant number of undistorted DBR periods after shrinkage the templates we sent to thermal annealing are made with 20 periods.
    
\begin{figure*}
    \includegraphics[width=\linewidth]{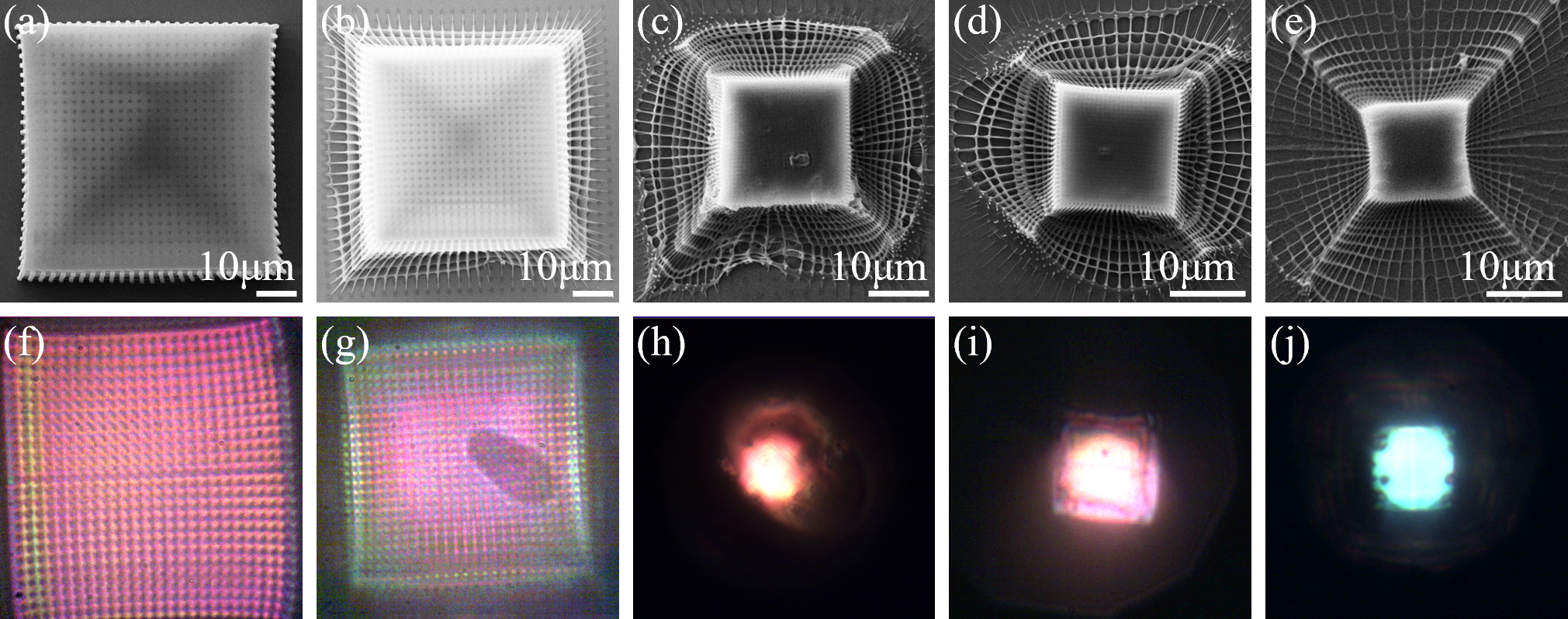}
    \caption{SEM images (top row: a-e) and corresponding reflection images under a white light source (bottom row: f-j) of pillar-supported DBR
    without shrinkage (a,~f) and with shrinkage times of 0 (b,~g), 4 (c,~h), 8 (d,~i), and 12 (e,~j) minutes respectively.}
    \label{fig:shrink}
\end{figure*}

    The SEM results are presented in Fig.~\ref{fig:shrink} with the corresponding white light illumination optical reflection.
    The optical reflection in Fig.~\ref{fig:shrink}(g) shows a pinkish hue when the structure slightly shrinks. 
    For Fig.~\ref{fig:shrink}(h), the light is mostly orange and red, indicating the DBR period is changed compared to the one in Fig.~\ref{fig:shrink}(g).
    The uneven light reflection can be attributed to the defect of the top layer.
    In Fig.~\ref{fig:shrink}(i), the centre of the optical image shows a mix of orange and yellow.
    Finally, after 12 min in Fig.~\ref{fig:shrink}(j), a bright centre appears in blue, indicating significant shrinkage of the period of the DBR template after a 12-minute annealing process.
    Those optical changes will be discussed with optical measurements in the later section.
    In addition, it can be seen from the SEM images that upper layers of the structure shrink by increasing amounts during the annealing process. 
    The lowest layer is anchored to the substrate by the pillars which act as stress relief. 
    However the stress release process extends to higher layers of the structure. 
    Even for the minimal shrinkage (Fig.~\ref{fig:shrink}(b)) 2-3 layers are involved while the number of 'stress release' layers increases as the shrinkage increases to a maximum of 9-10 in the longest annealed samples (Fig.~\ref{fig:shrink}(e)).  
    What is also clear is that this leads to a series of top layers that are unstressed and thus undistorted. 
    The SEM's also show that the in-plane dimension of the square DBR shrinks with annealing duration.
    
    We define the ratio of the dimension before and after the annealing process as the shrinkage ratio:
    \begin{equation}
        \varphi = \frac{original~size}{shrunk~size}
        \label{eq:ratio}
    \end{equation}

\begin{figure}
    \includegraphics[width=0.85\linewidth]{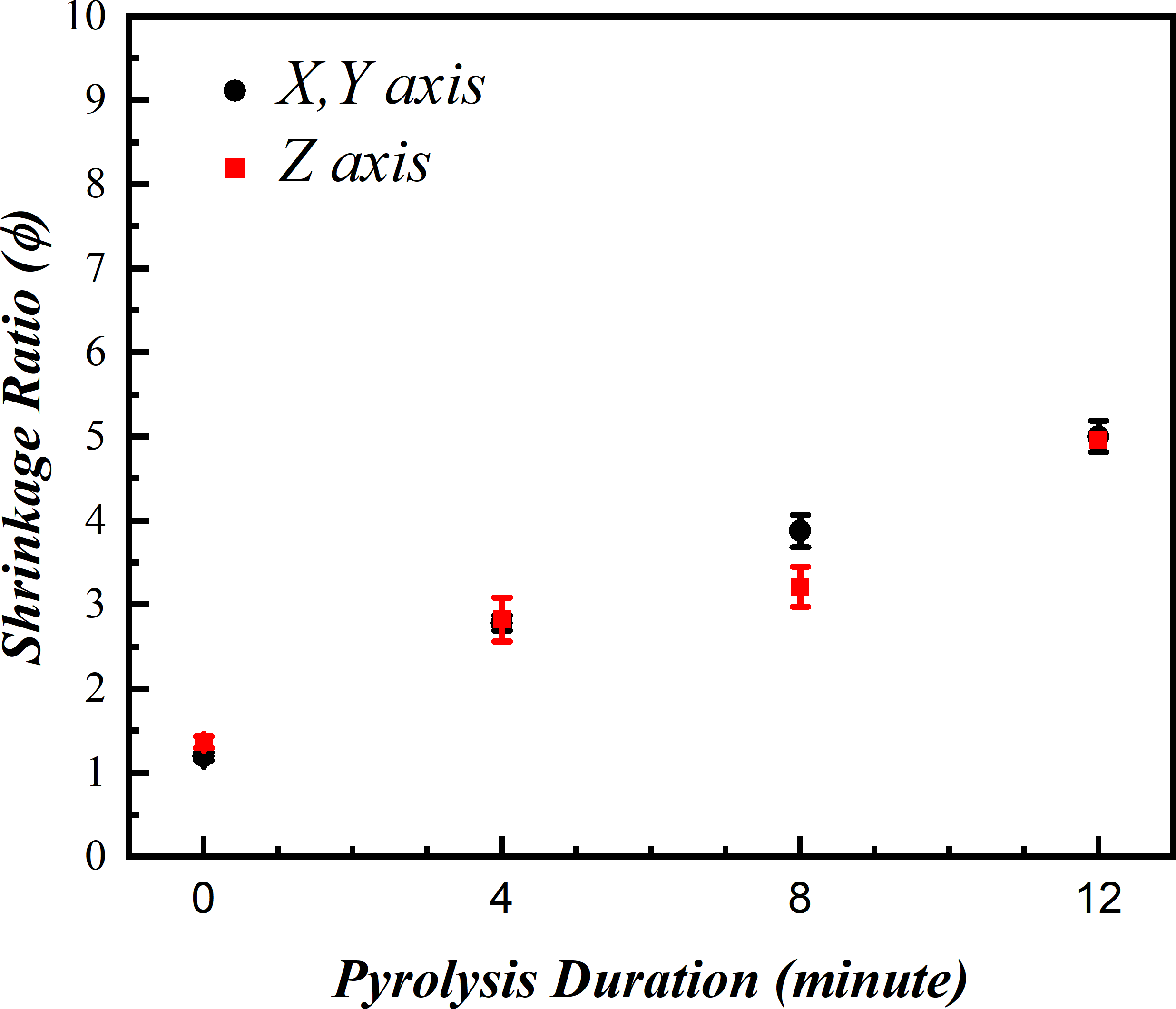}
    \caption{Shrinkage ratios with error bars as a function of different annealing times (See Fig.~\ref{fig:recipe}), from 0 minutes representing no annealing stage, to 12 minutes annealing time. The shrinkage ratio ascends to 5 times smaller than the original design, approximately.}
    \label{fig:ratio}
\end{figure}

    Assigning a shrinkage value of 1 to the original template we already see some shrinkage at 0 minutes of annealing ($\sim 1.5 $), during the heating up and cooling time.
    The shrinkage increases up to 5 as the annealing passes to 12 minutes, as shown in Fig.~\ref{fig:ratio}.

    These SEM's allow us to confirm shrinkage in-plane however the white light optical reflection results clearly show a shift toward the blue as shrinkage increases indicating that similar levels of shrinkage in the vertical dimensions is occurring.
    We can also measure the vertical period by using side view SEM's as in Fig.~\ref{fig:cross-section}.
    We now analyse the optical properties in detail in the following section.

\section{Results and discussion}

    \subsection{Optical characterisation and estimation}

\begin{figure*}
    \includegraphics[width=1.0\linewidth]{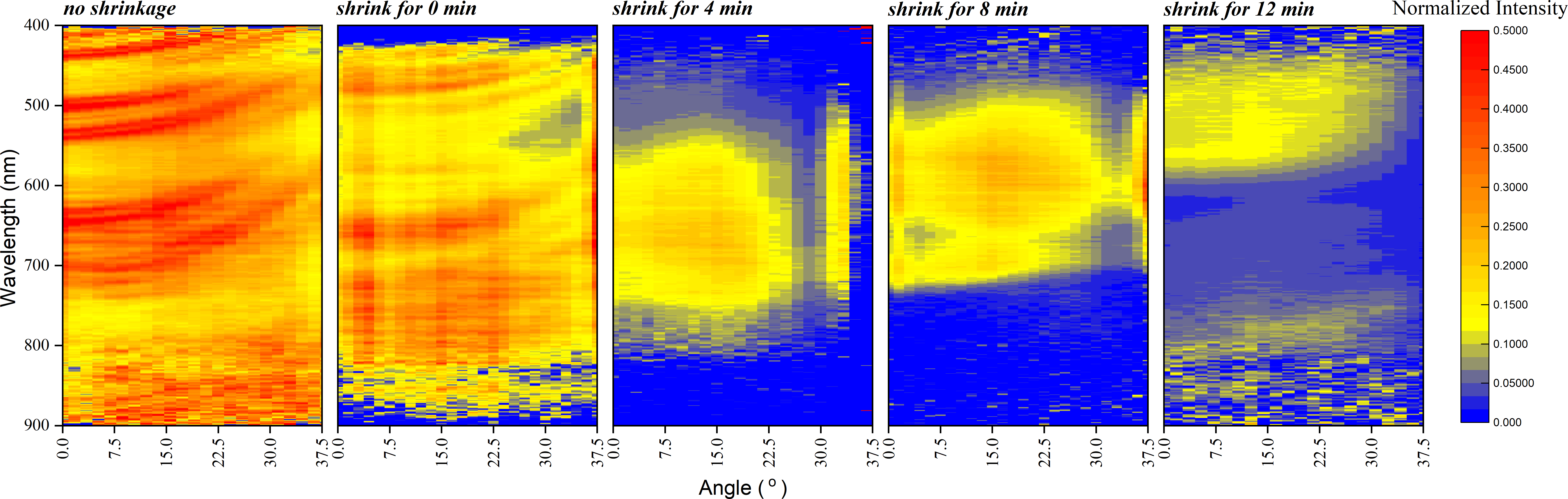}
    \caption{Measured angular reflection spectrum from 0° to 37.5°. Following the sequence from the original, 0, 4, 8 and 12 minutes. The results are normalized by mirror reflection.}
    \label{fig:banddiagram}
\end{figure*}

    The identification of the near infrared reflection seen in Fig.~\ref{fig:shrink} is facilitated through FIS.
    A white light source serves as the illumination in the visible/NIR and the spectrum of the reflected light is measured at various angles by scanning a collection fibre across the Fourier plane.
    The results are shown in Fig.~\ref{fig:banddiagram} using the four identical samples annealed for durations of 0, 4, 8, and 12 minutes as in Fig.~\ref{fig:shrink}.
    We see that as the templates contract, a noticeable shift in the bandgap is observed which we attribute to the reduced period of the DBR structure.
    We see also that most of the bands are blue shifted at higher angles indicative of Bragg reflection behaviour. 

    This suggests that the pillar-supported DBR can be approximated as a simple Bragg reflector.
    Bragg’s law, which is derived from the constructive interference of light waves reflecting off the different layers is given by the following relationship\cite{saleh2007:FundamentalsOfPhotonics,Joannopoulos:08:Book}:
    \begin{equation}
        m \lambda =2d, \ \ \ \ \ 
        m = 1,\ 2,\ 3,\ ...
        \label{eq:bragg}
    \end{equation}   
    where $m$ is an integer, $d$ represents the optical thickness of a single period, which at an incidence angle $\theta$ is:
    \begin{equation}
        d= n_1d_1\cos(\theta_1) + n_2d_2\cos(\theta_2) 
    \end{equation}
    where $\theta_{1/2}$ are the propagation directions in layers $d_{1/2}$ and $n_{1/2}$ their respective refractive indices. For normal incidence light, $\theta$ = 0, where the $\cos(\theta) = 1$. and for optical thickness $d$: 
    
        \begin{equation}
            d= n_1d_1 + n_2d_2 \label{eq:optic_thickness}
        \end{equation}
        
    We have estimated the values of $d_1$ and $d_2$ from side view SEM images with an example seen in Fig.~\ref{fig:cross-section}.
    We also estimate the supporting pillar diameters and separation and refractive index of the polymer and show these in Table~\ref{table:parameter}.
    Tables~\ref{table:higher-order} and ~\ref{table:band-position} then show estimated bandgap positions of fundamental and higher order bandgaps calculated from equations~\ref{eq:bragg} and \ref{eq:optic_thickness}, predicting where we expect to see reflection features at normal incidence. 

    \begin{table}
    \caption{\label{table:parameter}
        The summary of template parameters after different post-processing recipes. 
        \(d_1\) and \(d_2\) are the air and polymer thickness in micrometre. \(n_1\) and \(n_2\) is the refractive index of the material. \(n_2\) is increased after thermal post-processing due to solvent evaporation, and polymer carbonisation \cite{Liu2019_10.1038/s41467-019-12360-w}.
    }
    \begin{tblr}{
        row{1}={font=\boldmath\bfseries},
        hline{1,2,Z} = {1}{-}{}, 
        colspec={X[-1,c]X[-1,c]X[-1,c]X[-1,c]X[-1,c]X[-1,c]X[-1,c]},
        row{1}={valign=m},
        columns={halign=c},
    }
         \emph{Sample} & \(d_1\)(µm) & \(n_1\) & \(d_2\)(µm) & \(n_2\) & {\emph{Pillar}\\\emph{Diameter}(µm)} & \emph{Pillar Period}(µm) \\ 
         \emph{Original}         & 2.81  & 1 & 1.11  & 1.57 & 1.18 & 2.67 \\
         \emph{Shrink for 0 min}  & 1.92  & 1 & 0.96  & 1.57 & 1.05 & 2.32 \\
         \emph{Shrink for 4 min}  & 1.02  & 1 & 0.37  & 1.75 & 0.40 & 0.82 \\ 
         \emph{Shrink for 8 min}  & 0.82  & 1 & 0.40  & 1.75 & 0.38 & 0.80 \\ 
         \emph{Shrink for 12 min} & 0.54  & 1 & 0.25  & 1.75 & 0.26 & 0.56 \\
    \end{tblr}
\end{table}
    \begin{table}
    \caption{\label{table:higher-order}
        Estimated higher-order bandgap centre positions based on the given parameters for different post-processing durations from Table~\ref{table:parameter} using the Bragg's law approximation.
    }
    \begin{tblr}
    {
        row{1}={font=\bfseries},
        colspec={X[-1,l]X[-1,c]X[-1,l]X[-1,l]X[-1,l]X[-1,l]X[-1,l]},
        hline{6},
        hline{1,2,Z} = {1}{-}{}, 
        cell{even}{3-Z}={gray!50},
        cell{2,6}{1,2}={r=4}{m},
        cell{1}{3}={c=5}{c},
    }
    Template & Fundamental & Higher-order band \\
    Original & 9.10 $\mu m$ & 8$^{th}$ 		& 9$^{th}$ 			& 10$^{th}$ 	& 11$^{th}$ 	& 12$^{th}$ \\
    & 						& 1.13 $\mu m$ 	& 1.01 $\mu m$  	& 0.91 $\mu m$ 	& 0.83 $\mu m$  & 0.76 $\mu m$ \\
    & 						& 13$^{th}$ 	& 14$^{th}$ 		& 15$^{th}$ 	& 16$^{th}$ 	& 17$^{th}$ \\
    & 						& 0.70 $\mu m$	& 0.65 $\mu m$  	& 0.61 $\mu m$ 	& 0.57 $\mu m$ 	& 0.54 $\mu m$ \\
    Shrink for 0 min 	& 6.86 $\mu m$ 	& 8$^{th}$ 		& 9$^{th}$ 			& 10$^{th}$ 	& 11$^{th}$ 	& 12$^{th}$ \\
    & 									& 0.84 $\mu m$  & 0.76 $\mu m$  & 0.67 $\mu m$  & 0.62 $\mu m$  & 0.57 $\mu m$ \\
    & 									& 13$^{th}$ 	& 14$^{th}$ 		& 15$^{th}$ 	& 16$^{th}$ 	& 17$^{th}$ \\
    & 									& 0.53 $\mu m$  & 0.49 $\mu m$  & 0.46 $\mu m$  & 0.43 $\mu m$  & 0.40 $\mu m$     
    \end{tblr}
\end{table}
    \begin{table}
    \caption{\label{table:band-position}
        Estimated lower-order bandgap centre positions based on the given parameters for different post-processing durations from Table~\ref{table:parameter} using the Bragg's law approximation.
    }
    \begin{tblr}{
        row{1}={font=\boldmath\bfseries},
        colspec={X[l]X[-1,l]X[-1,l]X[-1,l]X[-1,l]X[-1,l]},
        hline{1,2,Z} = {1}{-}{}, 
    }
    Template & Fundamental  & $2^{nd}$ order & $3^{rd}$ order & $4^{th}$ order & $5^{th}$ order    \\
    \emph{Original} & 9.10 $\mu m$ & 4.55 $\mu m$ & 3.03 $\mu m$ & 2.28 $\mu m$ & 1.82 $\mu m$ \\
    \emph{Shrink\ for\ 0\ min} & 6.86 $\mu m$ & 3.43 $\mu m$ & 2.29 $\mu m$ & 1.71 $\mu m$ & 1.37 $\mu m$ \\
    \emph{Shrink\ for\ 4\ min} & 3.34 $\mu m$ & 1.67 $\mu m$ & 1.11 $\mu m$ & 0.84 $\mu m$ & 0.67 $\mu m$ \\
    \emph{Shrink\ for\ 8\ min} & 3.04 $\mu m$ & 1.52 $\mu m$ & 1.04 $\mu m$ & 0.75 $\mu m$ & 0.61 $\mu m$ \\
    \emph{Shrink\ for\ 12\ min} & 1.96 $\mu m$& 0.98 $\mu m$ & 0.65 $\mu m$ & 0.49 $\mu m$ & 0.39 $\mu m$ \\
    \end{tblr}
\end{table}
     
   This indicates that the fundamental bandgap wavelength for all five structures extends beyond the measurement range of the current FIS setup.
    The fundamental bandgap might be moved to a shorter wavelength by reducing the initial dimensions. However initial tests with thinner period samples have been unsuccessful due to vertical resolution of the writing process and uniformity of the shrinking across sample size >10$\mu$m, the minimum sample dimension for our FIS system\cite{10.1021/acsphotonics.9b00184}.
    As a result, we shifted our focus to the higher-order bands.
    The FIS results in Fig.~\ref{fig:banddiagram} for the original sample shows several higher order bands.
    If we compare the band positions at normal incidence to equation~\ref{eq:optic_thickness}, we can identify various high order bands in the NIR in Tables~\ref{table:higher-order} and \ref{table:band-position}.
    For instance in the original sample the 505 nm band is an $m=18$ band, the 650 nm band ($m=14$), the 700 nm band ($m=13$) and the weaker 900 nm band with $m=10$. However, bands seem to merge at shorter wavelengths, and notably, some are missing. Comparing this with the 0 minutes shrinkage sample, we see the $m=14$ band is predicted to move to 490 nm, and we see a band at 480 nm, and the $m=10$ band should move to 686 nm while the results show a nearby strong band centred around 670 nm. In addition the strong $9^{th}$ order band now appears in range around 740 nm and potentially a slightly blue shifted 8th order band is mixed with the $9^{th}$ order band creating a broad reflection feature.
    These blue shifts from predictions could be explained if we have underestimated the increase in the refractive index of the polymer, as measuring the exact refractive index of the polymer at different stages of the thermal treatment remains elusive, and we treat the 0-minute sample refractive index as unchanged (1.57) when calculating band positions.
    For 4-12 minute annealed samples, we assume the post-annealing 1.75 value.
    Normal incidence reflection in the FIS results supports the optical microscope reflection colours seen in Fig.~\ref{fig:shrink}.
    For the 4 and 8-minute  annealed templates the FIS identifies a $5^{th}$ order band moving from 670 $\pm$30 nm to 600 $\pm$ 30 nm. Although there is a hint of a separate 4th order band around 720 $\pm 20$ nm 
    For the 12-minute sample, the $5^{th}$ band is beyond the visible setup's range (at 0.39 $\mu$m), and only a weak reflection from the $4^{th}$ band is detected around 520 $\pm 30$ nm. There is also what appears to be the edge of a broad second order band appearing at the edge of the spectrometer range above 850 nm. 
    The bandgap structure away from normal shows the typical blue shifting expected from Bragg reflective structures. However, all bands observed in the 4-12 minute shrunk samples appear to be much weaker and more diffuse than expected. This may be due to uneven shrinkage through the structure and also to simplifications of our Bragg model, which ignores the 2-dimensional pillar support network.

\subsection{Numerical simulation}
            
\begin{figure}
    \includegraphics[width=1\linewidth]{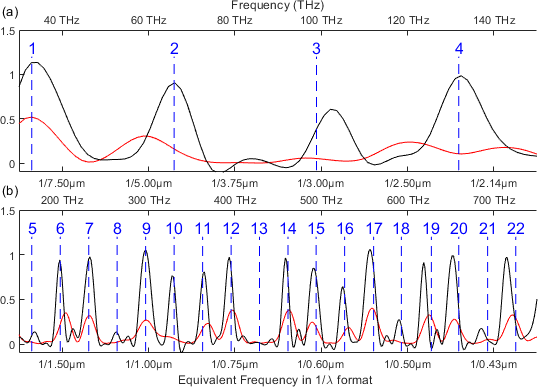}
    \caption{Full 3D FDTD simulations of DBR reflection coefficients plotted against frequency (upper linear axis) and equivalent wavelength (lower non-linear axis).
    Comparison between a simple DBR (black line) with a DBR and supporting pillars (red line).
    Resonances predicted by our simplest analysis (equation~\ref{eq:bragg}) are shown as blue dashed lines.
    The wavelength range extends to 10 $\mu m$ to observe the fundamental band, which is around 9.1 $\mu m$.
    The range is split into two plots (a) 2-10 $\mu$m and (b) 0.4-2 $\mu$m, allowing the detail of the higher-order short wavelength bands to be seen.}
    \label{fig:wo_pillar_FDTD}
\end{figure}

    To estimate the expected band positions better, we simulated the full structures using the FDTD method (using the Lumerical software package). 
    The angular resolved analysis of the 3D complex structure, including the support pillars, takes significant computing time, and initially, we look just at the simple 0-degree reflection over a range of wavelength out to 10 $\mu$m for unshrunk structures with and without pillars.
    Fig.~\ref{fig:wo_pillar_FDTD} illustrates that our simple analysis (see equation \ref{eq:bragg} and Tables~\ref{table:higher-order} and \ref{table:band-position}) predicts well the positions of low order bands and is only blue shifted slightly from the FDTD above band 14.
    Also we note that some bands are suppressed, notably 5$^{th}$, 8$^{th}$, 13$^{th}$, 18$^{th}$ and 21$^{st}$ bands.
    However our optical measurements (Fig.~\ref{fig:banddiagram}) on the original unshrunk sample seemingly contradict this as we see an apparently strong 13th order reflection. The disappearance of bands can be attributed to the non-quarter wavelength DBR design of our template. By introducing small changes in parameters $d_1$ and $d_2 \times n_2$ we can return this band and suppress neighbouring bands. When the pillars are introduced the band positions do not shift too much but the band reflection  coefficients are suppressed partly because the area of DBR is reduced with further subtle effects arising from pillar scattering. 
    
\begin{figure*}
    \subfloat[][0 min]{
        \includegraphics[width=0.5\linewidth]{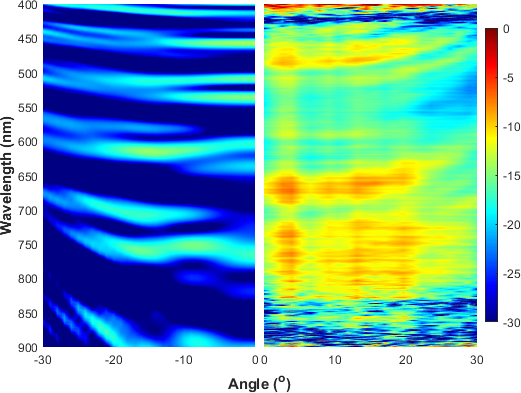}
        \label{fig:s0}
    }
    \subfloat[][4 min]{
        \includegraphics[width=0.5\linewidth]{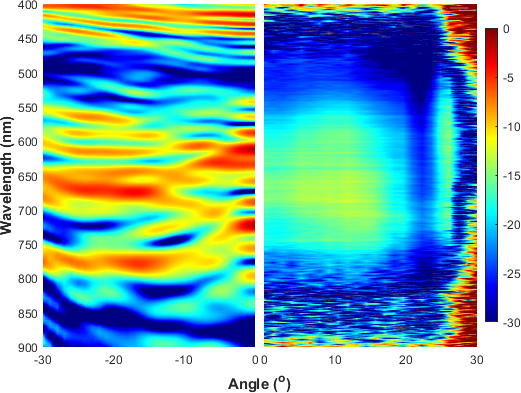}
        \label{fig:s4}
    }

    \subfloat[][8 min]{
        \includegraphics[width=0.5\linewidth]{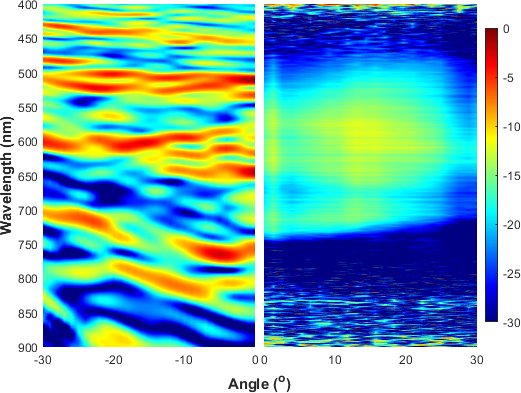}
        \label{fig:s8}
    }
    \subfloat[][12 min]{
        \includegraphics[width=0.5\linewidth]{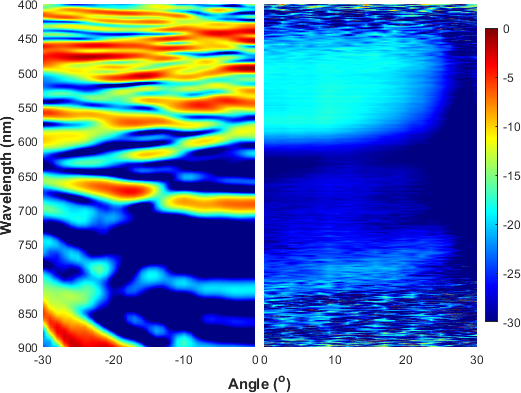}
        \label{fig:s12}
    }
    
    \caption{
        Comparison of normalized computational results (left) and experimental results (right) in decibels for the (a) 0, (b) 4, (c) 8, and (d) 12 min thermal annealing durations.
    }
    \label{fig:sim_vs_meas}
\end{figure*}

    We have made some 3D FDTD estimations of the angularly resolved scattering expected from our annealed and shrunk samples to help confirm the bandgap identification in our measurement results.
    Direct comparisons between smoothed FDTD simulations and experimental FIS results are shown in Fig.~\ref{fig:s0} to Fig.~\ref{fig:s12}.
    To better see weak bands we have changed the colour map scale to logarithmic.
    The FDTD results show band mixing effects between the 2D pillar array and the DBR structure making for a very complex bandstructure even after we apply our 2D smoothing to the results\cite{Origin2019}.
    
    In the figures (\ref{fig:s0} to ~\ref{fig:s12}) we reverse the angle axis of the simulations so that zero degrees appears at the centre to better compare with measured data. Simulation parameters used are taken from Table~\ref{table:parameter}.
    Despite the poor quality of reflection after shrinking we can still make some generic comparisons between simulations and optical measurements.
    
    In Fig.~\ref{fig:s0}, we see features associated with the 8$^{th}$, 9$^{th}$ and 10$^{th}$ (and possibly 11$^{th}$) bands spread over the range 600-850 nm. The 8$^{th}$ and 9$^{th}$ overlap well with simulation but 10$^{th}$ band identified earlier does not have a simulation counterpart. The logarithmic scaling has revealed weaker bands in the measurements that could correspond to 11$^{th}$, 12$^{th}$ or 13$^{th}$ bands which do appear in the simulations. However the pillar structures do seem to have disrupted the DBR bands making them split at higher angles. 

    For the structure subjected to a 4-minute post-process in Fig.~\ref{fig:s4}, the simulation shows some alignment with the experimental results. 
    The broad normal incidence 5$^{th}$ order band seen in the measurement centred around 650 nm is also seen in the FDTD simulations albeit split up again possibly due to interference with pillar induced bandstructure. 
    
    Similarly the edge of a fourth order band feature appears at short wavelength (400-450 nm) in both measurement and simulation. 
    
    In Fig.~\ref{fig:s8}, the simulated bands appear cluttered . Nonetheless, it's evident that the most intense reflection is concentrated between 550 nm and 720 nm, aligning with the experimental findings. 
    In the measured data, the band's absence around 800 nm can be attributed to the diminished reflection intensity and maybe from layer spacing varying with depth, which we will investigate in future work.
        
    In Fig.~\ref{fig:s12}, the simulated band predominantly centres around 500 nm, aligning with the faint 4$^{th}$ order reflection observed in the experimental data. The third order band appears as a strong feature at 650-700nm but only weakly in the measurement while the second order band is clearly at longer wavelengths than predicted in table~\ref{table:band-position}.


\section{Discussion and Conclusion}
    In this study, our objective was to establish a consistent and predictable methodology for thermal post-processing, enabling the controlled shrinkage of arbitrary 3D PhC by annealing polymer templates.
    Here, we restrict ourselves to the simplest PhC structure that of a one dimensional DBR albeit stabilised by a pillar array between layers fabricated using 2PP-DLW. A commercial RTA was then used to anneal the structure evapourating unpolymerised material thus shrinking the structure. 
   
    Structures were characterised by optical and electron microscopes monitoring overall the structural quality and determining parameters for computational predictions.
    Subsequently, angular-resolved reflection measurements were conducted using our custom-built FIS system. The reflected light was gathered and represented as colour map band diagrams covering scattering angles up to 40 degrees and spanning the 400-900 nm wavelength band.

    The band positions were predicted from structural parameters determined in SEM measurements using simple Bragg theory and commercial simulation software for TMM, and full 3D FDTD. In the latter case we were able to included the pillar arrays.

    The first conclusion is that the fundamental band gap in all our samples appeared at too long wavelengths to be measured in our FIS system. However the results do show that bands as low as second order begin to appear in the 12 minute shrinkage sample where a shrinkage factor of up to 5 was measured. 

    However, we were able to identify higher order band features that tracked across the visible region leading to clear blue shifting of reflections in the optical microscope pictures. At first sight we thought a single low order (5$^{th}$ order band) might be responsible but in fact the shrinkage was too great moving the 5$^{th}$ order band out of the visible into the UV revealing a visible fourth order band in the final 12 minute shrinkage sample.  
    
    Before shrinkage several higher order bands were identified along with notable missing bands which we ascribe to the DBR having unbalanced optical thickness between air and polymer layers. After shrinking the bands broaden partly because they are lower order and possibly widenened by non-uniform layer thickness and distortion post-thermal shrinkage. 
    There will be a gradient in layer thickness particularly in the lowest few layers where stretching occurred due to anchorage at the substrate.
    It is clear from figure~\ref{fig:shrink} that the first pillar layer acts as a stress relieving layer (~1$\mu$m tall), but higher layers are stretched as well.
    In the highest shrinkage sample (Fig.~\ref{fig:shrink}(e)) stretching is clearly extending up to layer 9-10 of the 20 layer structure.
     
    Estimating the effective layer thickness for simulations to fit the experimental results was challenging but overall similar bands could be identified linking all the simulation methods to the measured reflection data. 
    
    In summary, this research has effectively showcased the potential of employing thermal shrinkage of polymer-based DBR structures as a reliable technique for bandgap modifications. 
    Our study reveals that a uniform shrinkage ratio of up to five times is achievable, paving the way for significant alterations in 3D PhC bands, thanks to the scaling properties inherent in Maxwell's equations. As the period of the DBRs diminishes, we observe pronounced blue shift in higher-order bandgaps within the visible range.  
    This result is evident in the optical reflection pictures of the shrunk DBR samples (Fig.~\ref{fig:shrink}), where iridescent reflection colours are clearly seen shifting towards the blue with increasing shrinkage. 
    Despite challenges we have succeeded in aligning experimental data roughly with computational predictions 
    
    Ongoing efforts are directed towards minimising fabrication discrepancies and refining simulation fits. Furthermore, these templates have sparked the idea of employing a buffer structure when subjecting any arbitrary 3D templates to thermal post-processing. This concept will be a candidate when considering thermal post-processing on polymer-based PhC's in future research.
    


\medskip
\textbf{Acknowledgements} \par 
This work was supported by the Engineering and Physical Sciences Research Council (EPSRC) grants EP/N00762X/1, EP/V040030/1, EP/Y003551/1, and EP/Y016440/1.

\bibliographystyle{iopart-num}
\bibliography{main.bib}

\end{document}